\documentclass{elsart}
\begin{document}
\begin{center}
 {\large\bf Jaynes-Cummings model without rotating wave approximation.
            Asymptotics of eigenvalues.}
\end{center}
\begin{center}
 {\it E.A.Tur\\ \medskip {
Department of Higher Mathematics\\
St--Petersburg State Institute of\\
Fine Mechanics and Optics (Technical University)\\
Sablinskaya 14, 197101 St--Petersburg, Russia}\\ 
 {\sf E-mail:} Teduard@cards.lanck.net}
\end{center}
\begin{quote}
{\footnotesize
{\bf Abstract.} In this paper the perturbation theory with the frequency of
                transition in atom as perturbation parameter is constructed.
                The estimation of the reminder term of series of this
                perturbation theory is given. With the help of this
                perturbation theory we have found an exact asymptotics of
                eigenvalues of complete hamiltonian in the
                limit of high quantum numbers. It is shown that the
                counter-rotating terms keep a leading term but
                absolutely change a second term of this asymptotic.}
\end{quote}

\bigskip
{\bf 1. Introduction.}
\bigskip

The Jaynes-Cummings model without rotating wave approximation (RWA) is the
elementary model describing an interaction of atom with a field. But despite
of this it can not be solved exactly. This model without the RWA was
considered by different methods in works~[1-7]. The hamiltonian of this model
has the form

\begin{equation}
\label{1}
  {\bf H}={\bf H}_0+g\,{\bf V}=\omega_0\,{\bf\sigma}_0+\omega\:
  {\bf a}^+\,{\bf a}+g\,{\bf\sigma}_1\,
  (\,{\bf a}+{\bf a}^+\,)\,,
\end{equation}
where ${\bf a}$ and ${\bf a}^+$ are the photon creation and annihilation
operators, $g$ is the coupling constant, $\omega$ and $\omega_0$ are the
frequencies of mode and atomic transition respectively,
${\bf\sigma}_0$ and ${\bf\sigma}_1$ are the $2\times2$ matrices of form
$$
{\bf\sigma}_0=\left(\begin{array}{cc} 2 & 0\\0 & 1\end{array}\right)
\,,\quad
{\bf\sigma}_1=\left(\begin{array}{cc} 0 & 1\\1 & 0\end{array}\right)
$$

It is well known that the RWA formulas for eigenvalues take into account only
zero and first order of the perturbation theory on the coupling constant $g$.
Therefore they are valid only at small relative coupling constant $g/\omega$
and sufficiently small quantum index. More precisely, the validity of the RWA
formulas for eigenvalues is defined by the condition $g\sqrt n/\omega\ll1$.
In the case of exact resonance ($\omega=\omega_0$) the expression $g\sqrt n$
defines the splitting of eigenvalues. In optics $g/\omega\ll1$. Hence,
unique opportunity to leave for limits of the RWA is the consideration of
the highly exited states with suffisiently large quantum index $n$. That is
the RWA loses force at sufficiently large energies of a field mode. How the
eigenvalues of the hamiltonian~(\ref{1}) and the splitting of them behave at
the large quantum indexes? In the present paper we shall answer on this
question by constructing the perturbation theory on the parameter $\omega_0$,
which enters linearly in the hamiltonian~(\ref{1}). We shall show that this
perturbation theory well describes not only an eigenvalues at
$\omega_0/\omega\ll1$ but also the highly laying eigenvalues at arbitrary
$g$ and $\omega_0\le k\,\omega$, where $k=\sqrt3/(2\pi)\simeq0.23$ .
We also give an estimation of the reminder term of series and find two first
terms of asymptotic of eigenvalues on quantum index. It is interesting that
the second term of this asymptotic is qualitatively differed from the
corresponding term in the RWA. This difference leads to the fact that the
splitting of eigenvalues
vanishes at the large quantum numbers, unlike the RWA case, when the
splitting infinitely increases.

We remain open the question about the validity of the Jaynes-Cummings model
itself at the limit of large average energy of a field mode, because then the
manyphoton transitions between other levels become important. Nevertheless,
we can hope, that the resonant levels give the basic contribution in atomic
dynamics even for large average energy of a field mode.

\bigskip
{\bf 2. Perturbation theory on the parameter $\omega_0$.}
\bigskip

In work~\cite{6} we have shown that the hamiltonian of model~(\ref{1}) can be
represented in invariant subspaces by two Jacobi matrices of form
$$
  {\bf H}_1=\left(\begin{array}{cccccc} \omega_0&g\sqrt{1}&0&0&0&\ldots\\
                      g\sqrt{1}&2\omega_0+\omega&g\sqrt{2}&0&0&\ldots\\
                      0&g\sqrt{2}&\omega_0+2\omega&g\sqrt{3}&0&\ldots\\
                      0&0&g\sqrt{3}&2\omega_0+3\omega&g\sqrt{4}&\ldots\\
                      0&0&0&g\sqrt{4}&\omega_0+4\omega&\ldots\\
                      \ldots&\ldots&\ldots&\ldots&\ldots&\ldots
  \end{array}\right)
$$
\vspace{0.5cm}
$$
  {\bf H}_2=\left(\begin{array}{cccccc} 2\omega_0&g\sqrt{1}&0&0&0&\ldots\\
                      g\sqrt{1}&\omega_0+\omega&g\sqrt{2}&0&0&\ldots\\
                      0&g\sqrt{2}&2\omega_0+2\omega&g\sqrt{3}&0&\ldots\\
                      0&0&g\sqrt{3}&\omega_0+3\omega&g\sqrt{4}&\ldots\\
                      0&0&0&g\sqrt{4}&2\omega_0+4\omega&\ldots\\
                      \ldots&\ldots&\ldots&\ldots&\ldots&\ldots
  \end{array}\right)
$$

Let us present the operators ${\bf H}_1$ and ${\bf H}_2$ as
\begin{equation}
\label{oper}
{\bf H}_1={\bf A}_0+\omega_0\,{\bf P}_1\,,\quad
{\bf H}_2={\bf A}_0+\omega_0\,{\bf P}_2\,,
\end{equation}
where ${\bf A}_0$ is the unbounded main operator without periodic modulation
of main diagonal, ${\bf P}_1$ and ${\bf P}_2$ are the diagonal projectors
\begin{equation}
\label{2}
{\bf A}_0=\left(\begin{array}{cccccc} \omega_0&g\sqrt{1}&0&0&0&\ldots\\
                      g\sqrt{1}&\omega_0+\omega&g\sqrt{2}&0&0&\ldots\\
                      0&g\sqrt{2}&\omega_0+2\omega&g\sqrt{3}&0&\ldots\\
                      0&0&g\sqrt{3}&\omega_0+3\omega&g\sqrt{4}&\ldots\\
                      0&0&0&g\sqrt{4}&\omega_0+4\omega&\ldots\\
                      \ldots&\ldots&\ldots&\ldots&\ldots&\ldots
  \end{array}\right)
\end{equation}
\begin{equation}
\label{3}
{\bf P}_1=\left(\begin{array}{cccccc} 0&0&0&0&0&\ldots\\
                      0&1&0&0&0&\ldots\\
                      0&0&0&0&0&\ldots\\
                      0&0&0&1&0&\ldots\\
                      0&0&0&0&0&\ldots\\
                      \ldots&\ldots&\ldots&\ldots&\ldots&\ldots
  \end{array}\right)\:,\quad
{\bf P}_2=\left(\begin{array}{cccccc} 1&0&0&0&0&\ldots\\
                      0&0&0&0&0&\ldots\\
                      0&0&1&0&0&\ldots\\
                      0&0&0&0&0&\ldots\\
                      0&0&0&0&1&\ldots\\
                      \ldots&\ldots&\ldots&\ldots&\ldots&\ldots
  \end{array}\right)
\end{equation}

The operator ${\bf A}_0$ is the hamiltonian of shifted oscillator. It can be
diagonalized with the help of Bogolubov's transformation. Its eigenvalues and
eigenvectors have the form
\begin{equation}
\label{4}
{\bf A}_0\,|a_m\rangle=\lambda_m\,|a_m\rangle\,,\quad
\lambda_m=\omega_0+m\omega-g^2/\omega\,,\quad
|a_m\rangle=\sum_{n=0}^\infty P^{(m)}_n\,|e_n\rangle\,,\quad
\langle a_m|a_n\rangle=\delta_{m,n}\,,
\end{equation}
where $\{|e_n\rangle\}$ is the basis of matrix representation~(\ref{2}),
$P^{(m)}_n$ are defined by Feynman-Schwinger's formulas~\cite{8,9}
\begin{equation}
\label{5}
P^{(m)}_n=\exp\{-g^2/2\omega^2\}\,\sqrt{\frac{n!}{m!}}\,\left(\frac{g}{\omega}
\right)^{m-n}\,L_n^{m-n}(g^2/\omega^2)
\end{equation}

Here $L_n^{s}$ are generalized Chebyshev-Laguerre's polynomials
$$
L_n^{s}(x)=\frac{(n+s)!}{n!}\,\sum_{i=0}^n C_n^i\,(-1)^i\,\frac{x^i}{(i+s)!}
\,,\quad C_n^i=\frac{n!}{i!\,(n-i)!}
$$

It is easy to veryfy that the expression~(\ref{5}) can be presented also in
the form of contour integral
\begin{equation}
\label{contour}
P^{(m)}_n=
\exp\{-g^2/2\omega^2\}\,\sqrt{\frac{m!}{n!}}\,\left(\frac{g}{\omega}
\right)^{n-m}\,\frac{1}{2\pi i}\,
\oint\limits_C x^{m-1}\left(\frac{1}{x}-1\right)^n
\exp\left\{\frac{g^2}{\omega^2}\,\frac{1}{x}\right\}dx
\end{equation}
where $C$ is the circle of unit radius with the centre in the origin of
coordinates of a complex plane $x$. This expression we will use further.

From~(\ref{oper}) it follows that the operators ${\bf H}_1$ and ${\bf H}_2$
depend linearly on $\omega_0$. If we know the solution of spectral problem
for the operator ${\bf A}_0$, we can build the perturbation theory on the
parameter $\omega_0$. Let us find the matrix form of the operators ${\bf P}_1$
and ${\bf P}_2$ in the basis of the operator ${\bf A}_0$ eigenvectors. Let
${\bf U}(g)$ be the orthogonal transformation from $|e_m\rangle$ at
$|a_m\rangle$
\begin{equation}
\label{6}
|a_m\rangle={\bf U}(g)\,|e_m\rangle\,,\quad
\langle e_k|\,{\bf U}(g)\,|e_m\rangle=P^{(m)}_k\,,\quad
{\bf U}^T\,{\bf U}={\bf E}
\end{equation}

Using~(\ref{3}),(\ref{4}) and~(\ref{6}), we have
\begin{equation}
\label{7}
P^{(1)}_{k,m}\equiv\langle a_k|{\bf P}_1|a_m\rangle=
\langle e_k|{\bf U}^T\,{\bf P}_1\,{\bf U}|e_m\rangle=
\langle e_k|{\bf P}^{(1)}|e_m\rangle=
\sum_{n-odd}P^{(k)}_n\,P^{(m)}_n
\end{equation}
\begin{equation}
\label{8}
P^{(2)}_{k,m}\equiv\langle a_k|{\bf P}_2|a_m\rangle=
\langle e_k|{\bf U}^T\,{\bf P}_2\,{\bf U}|e_m\rangle=
\langle e_k|{\bf P}^{(2)}|e_m\rangle=
\sum_{n-even}P^{(k)}_n\,P^{(m)}_n\,,
\end{equation}
where ${\bf P}^{(1)}={\bf U}^T\,{\bf P}_1\,{\bf U}$ and
${\bf P}^{(2)}={\bf U}^T\,{\bf P}_2\,{\bf U}$ are the transformed projectors.

Let us consider for example the sum~(\ref{7}). Using the
representation~(\ref{contour}) and summarizing on odd values $n$, we come to
the formula
$$
P^{(1)}_{k,m}=
\exp\{-g^2/\omega^2\}\,\sqrt{k!\,m!}\,\left(\frac{g}{\omega}\right)^{-m-k}\,
\cdot
$$
$$
\cdot\frac{1}{(2\pi i)^2}\,\oint\limits_C\oint\limits_C (x)^{m-1}\,(x')^{k-1}\,
sh\left\{\frac{g^2}{\omega^2}\left(\frac{1}{x}-1\right)\left(\frac{1}{x'}
-1\right)\right\}
\exp\left\{\frac{g^2}{\omega^2}\left(\frac{1}{x}+\frac{1}{x'}\right)
\right\}dx\,dx'
$$

The contour integrals in this expression can be calculated consistently with
the help of residues. As a result, we obtain the following expression for
$P^{(1)}_{k,m}$
\begin{equation}
\label{9}
P^{(1)}_{k,m}=\frac{1}{2}\,\delta_{k,m}-\frac{(-1)^k}{2}\,
\exp\left\{-\,\frac{2g^2}{\omega^2}\right\}\sqrt{\frac{m!}{k!}}\left(\frac{2g}
{\omega}\right)^{m-k}\,\sum_{i=0}^k C_k^i\,(-1)^i\,\frac{(4g^2/\omega^2)^i}
{(i+m-k)!}
\end{equation}

Similarly, one can obtain and the expression for $P^{(2)}_{k,m}$, defined by
the sum~(\ref{8})
\begin{equation}
\label{10}
P^{(2)}_{k,m}=\frac{1}{2}\,\delta_{k,m}+\frac{(-1)^k}{2}\,
\exp\left\{-\,\frac{2g^2}{\omega^2}\right\}\sqrt{\frac{m!}{k!}}\left(\frac{2g}
{\omega}\right)^{m-k}\,\sum_{i=0}^k C_k^i\,(-1)^i\,\frac{(4g^2/\omega^2)^i}
{(i+m-k)!}
\end{equation}

Comparing~(\ref{9}) and~(\ref{10}) with~(\ref{5}), we have
\begin{equation}
\label{11}
P^{(1)}_{k,m}=\frac{1}{2}\,\delta_{k,m}-\frac{(-1)^k}{2}\,P^{(m)}_k(2g)
\end{equation}
\begin{equation}
\label{12}
P^{(2)}_{k,m}=\frac{1}{2}\,\delta_{k,m}+\frac{(-1)^k}{2}\,P^{(m)}_k(2g)\,,
\end{equation}
or in the operator form
\begin{equation}
\label{13}
{\bf P}^{(1)}={\bf U}^T(g)\,{\bf P}_1\,{\bf U}(g)=
\frac{1}{2}\,\left({\bf E}-{\bf B}\,{\bf U}(2g)\right)
\end{equation}
\begin{equation}
\label{14}
{\bf P}^{(2)}={\bf U}^T(g)\,{\bf P}_2\,{\bf U}(g)=
\frac{1}{2}\,\left({\bf E}+{\bf B}\,{\bf U}(2g)\right)\,,
\end{equation}
where ${\bf B}$ is the diagonal matrix with elements
$B_{m,k}=(-1)^k\,\delta_{m,k}$ . Let us note that the matrices ${\bf B}$ and
${\bf U}$ satisfy to the identity
$$
[{\bf B}\,{\bf U}]^2={\bf E}
$$

The formulaes~(\ref{11}),(\ref{12}) ( or~(\ref{9}),(\ref{10}) ) allow to write
at once the approximated expressions for eigenvalues taking into account only
zero and first orders of the perturbation theory on $\omega_0$.
The first order correcton to an eigenvalues is defined by diagonal elements
of perturbation. Taking into account the formula~(\ref{4}) for the eigenvalues
of the operator ${\bf A}_0$ and the expressions~(\ref{11}),~(\ref{12}),
~(\ref{5}) (at $k=m$), we obtain the following approximated formulaes for
eigenvalues $\lambda_m^{(1)}$ and $\lambda_m^{(2)}$ of the operators
${\bf H}_1$ and ${\bf H}_2$ respectively
$$
\lambda_m^{(1)}\simeq 3\omega_0/2+m\omega-g^2/\omega-
\frac{(-1)^m\,\omega_0}{2}\,
\exp\left\{-\,\frac{2g^2}{\omega^2}\right\}L_m(4g^2/\omega^2)
$$
$$
\lambda_m^{(2)}\simeq 3\omega_0/2+m\omega-g^2/\omega+
\frac{(-1)^m\,\omega_0}{2}\,
\exp\left\{-\,\frac{2g^2}{\omega^2}\right\}L_m(4g^2/\omega^2)
$$

This formulaes was obtained in work~\cite{5} with the help of a so-called
"operator method".

Let us consider now the constructed perturbation theory series in more detail.
We shall show that the two first term of this series give an exact asymptotic
of an eigenvalues $\lambda_m^{(1)}$ and $\lambda_m^{(2)}$ at large quantum
index $m$.

\bigskip
{\bf 3. Asymptotic of eigenvalues.}
\bigskip

Let us consider, for example, an eigenvalues $\lambda_m^{(2)}$ of the
operator ${\bf H}_2$. The proof of the appropriate formulas for
$\lambda_m^{(1)}$ is completely similarly. In what follows for brevity we
shall omit the top
index $(2)$ at eigenvalues and write $\lambda_m$ instead of $\lambda_m^{(2)}$.
The perturbation theory series for exact eigenvalue $\lambda_m$ has the form
\begin{equation}
\label{15}
\lambda_m=\sum_{k=0}^\infty\lambda_m^{(k)}\,,\quad \lambda_m^{(k)}\sim
(\omega_0)^k
\end{equation}

General expression for $\lambda^{(k)}_m$, in case when an operator depends
linearly on the perturbation parametr and an eigenvalues are not degenerate
(here, due to the simplicity of Jacobi matrix spectrum), has the
form~\cite{10}
\begin{equation}
\label{16}
\lambda^{(k)}_m=\frac{(-\omega_0)^k}{k}\sum_{\textstyle{
n_1+\ldots+n_k=k-1 \atop n_i\ge0}}\mbox{tr}
\left[{\bf P}\,{\bf S}_m^{n_1}\ldots{\bf P}\,{\bf S}_m^{n_k}\right]\,,\quad
k\ge1\,,
\end{equation}
where
\begin{equation}
\label{17}
{\bf S}_m^0\equiv-|a_m\rangle\langle a_m|\,,\qquad{\bf S}^n_m=\sum_{i\ne m}
\frac{|a_i\rangle\langle a_i|}{\left(\lambda_i^{(0)}-\lambda_m^{(0)}
\right)^n}=\frac{1}
{\omega^n}\sum_{i\ne m}\frac{|a_i\rangle\langle a_i|}{(i-m)^n}\,,\quad
n\ge1
\end{equation}

Here, we have omitted as well as above the top index $(2)$ at the perturbation
operator ${\bf P}^{(2)}$ and used the formula~(\ref{4}) for the unperturbed
eigenvalues $\lambda_m^{(0)}$.

We have found already that
\begin{equation}
\label{18}
\lambda_m^{(0)}=m\omega+\omega_0-g^2/\omega
\end{equation}
\begin{equation}
\label{19}
\lambda_m^{(1)}=\omega_0/2+\frac{(-1)^m\,\omega_0}{2}\,
\exp\left\{-\,\frac{2g^2}{\omega^2}\right\}L_m(4g^2/\omega^2)=
\omega_0/2+O(m^{-1/4})\,,\quad m\to\infty
\end{equation}
Here, we have used the asymptotic of Chebyshev-Laguerre's polynomials (see,
for example,~\cite{11}).

Let us consider the second order correction $\lambda^{(2)}_m$ which is
defined by
$$
\lambda_m^{(2)}=\omega_0^2\,\sum_{k\ne m}\frac{|P_{k,m}|^2}{\lambda_m^{(0)}-
\lambda_k^{(0)}}
$$
According to~(\ref{12}) and~(\ref{18}), this expression can be presented in
the form
$$
\lambda_m^{(2)}=\frac{\omega_0^2}{4\omega}\,
\sum_{k\ne m}\frac{[P^{(m)}_k(2g)]^2}{m-k}
$$
The behaviour of this expression as $m\to\infty$ is defined by the behaviour
of sum
\begin{equation}
\label{20}
t_m=\sum_{k\ne m}\frac{[P^{(m)}_k(2g)]^2}{m-k}
\end{equation}
Let us show that $t_m\to0$ as $m\to\infty$. For this purpose, let us
transform~(\ref{20}) to the form
$$
t_m=\sum_{n=1}^\infty\frac{C_{m,n}}{n}\,,
$$
where the transformation matrix $C_{m,n}$ is defined as follows
\begin{equation}
\label{21}
C_{m,n}=\left\{\begin{array}{r}
[P^{(m)}_{m-n}(2g)]^2-[P^{(m)}_{n+m}(2g)]^2\,,\quad n\le m\\
-[P^{(m)}_{n+m}(2g)]^2\,,\quad n>m
\end{array}
\right.
\end{equation}

The condition $t_m\to0$ follows from $\frac{\textstyle 1}{\textstyle n}\to0$,
if and only if the transformation $C_{m,n}$ satisfies to the
following conditions (~\cite{12}, Theorem 4)
\begin{equation}
\label{Hardy}
\begin{array}{l}
\displaystyle
1.\:\:\sum_n|C_{m,n}|<H\,,\quad\mbox{where\,\,$H$\,\,does not depend of $m$}\\
2.\:\:\lim\limits_{m\to\infty} C_{m,n}=0\,,\quad\mbox{for\,\,arbitrary\,\,$n$}
\end{array}
\end{equation}

Then $C_{m,n}$ is the regular transformation.
Let us prove the first condition. Taking into account~(\ref{21}), we have
$$
\sum_n|C_{m,n}|=\sum_{n=1}^m
\left[[P^{(m)}_{m-n}(2g)]^2-[P^{(m)}_{n+m}(2g)]^2\right]+
\sum_{n=m+1}^\infty[P^{(m)}_{n+m}(2g)]^2<
$$
$$
<\left[\sum_{n=0}^\infty[P^{(m)}_{n}(2g)]^2\right]-
[P^{(m)}_{m}(2g)]^2
$$

But since the values $P^{(m)}_{n}(2g)$ are the matrix elements of the
orthogonal transformation ${\bf U}(2g)$, the sum in square brackets is equel
to unit identically. The diagonal matrix element $P^{(m)}_{m}(2g)$, due to
~(\ref{5}), equals to
$$
P^{(m)}_{m}(2g)=\exp\left\{-\,\frac{2g^2}{\omega^2}\right\}L_m(4g^2/\omega^2)\,,
$$
and due to the asymptotic of Chebyshev-Laguerre's polynomials~\cite{11}, tends
to zero as $m\to\infty$. Therefore, we have
$$
\sum_n|C_{m,n}|<1-\delta_m\,,\quad\mbox{где}\:\:\delta_m\to0\:\:\mbox{as}\:\:
m\to\infty
$$
And hence, the condition 1 in~(\ref{Hardy}) is fulfilled.

Let us check now the validity of the second condition in~(\ref{Hardy}).
For this purpose, due to~(\ref{21}), it is necessary to consider the diagonal
asymptotic of the non-diagonal matrix elements of the transformation
${\bf U}(2g)$. From~(\ref{5}), we have
$$
P^{(m)}_{m-n}(2g)=
\exp\left\{-\,\frac{2g^2}{\omega^2}\right\}\sqrt{\frac{(m-n)!}{m!}}
\left(\frac{2g}{\omega}\right)^{n}\,L_{m-n}^{n}(4g^2/\omega^2)
$$
Using the asymptotic of generalized Chebyshev-Laguerre's
polynomials~\cite{11}
$$
L_{n}^{s}(x)=\pi^{-1/2}\,n^{s/2-1/4}\,x^{-s/2-1/4}\,e^{x/2}\left\{
\cos(2\sqrt{nx}-s\pi/2-\pi/4)+O(n^{-1/2})\right\}\,,\:n\to\infty\,,
$$
we obtain
$$
P^{(m)}_{m-n}(2g)\sim\frac{1}{m^{1/4}}\,,\:\:m\to\infty
\,,\quad\mbox{for\,\,arbitrary\,\,$n$}
$$
Due to the symmetry of perturbation matrix $P_{k,m}$, we obtain at once the
same asymptotic and for $P^{(m)}_{m+n}(2g)$
$$
P^{(m)}_{m+n}(2g)\sim\frac{1}{m^{1/4}}\,,\:\:m\to\infty
\,,\quad\mbox{for\,\,arbitrary\,\,$n$}
$$

According to~(\ref{21}), it follows that the condition 2 is also fulfilled.
Therefore due to the above theorem, $t_m$ and hence $\lambda^{(2)}_m$ tend to
zero as $m\to\infty$ :
\begin{equation}
\label{22}
\lambda^{(2)}_m\to0\,,\quad m\to\infty
\end{equation}

Let us consider the third order correction $\lambda_m^{(3)}$ to the
eigenvalue $\lambda_m$. From~(\ref{16}) and~(\ref{17}) it follows that
$\lambda_m^{(3)}$ is defined by expression
$$
\lambda_m^{(3)}=\frac{\omega_0^3}{\omega^2}\left[\sum_{i,j\ne m}\frac{
P_{m,i}\,P_{i,j}\,P_{j,m}}{(i-m)(j-m)}-P_{m,m}\,\sum_{i\ne m}\frac{
|P_{m,i}|^2}{(i-m)^2}\right]
$$
Using~(\ref{12}), we have
\begin{equation}
\label{23}
\left|\lambda_m^{(3)}\right|\le\frac{(\omega_0/2)^3}{\omega^2}
\left[\sum_{i,j\ne m}\frac{
|P^{(i)}_m(2g)|\,|P^{(j)}_i(2g)|\,|P^{(m)}_j(2g)|}{|i-m|\,|j-m|}+
\left|1+(-1)^m\,P^{(m)}_m(2g)\right|\,
\sum_{i\ne m}\frac{|P^{(i)}_m(2g)|^2}{(i-m)^2}\right]
\end{equation}

Let us consider the first term in square brackets. Let us aplay Cauchy's
unequality to the sum on $j$ in this composed
$$
\sum_{j\ne m}\frac{|P^{(j)}_i(2g)|\,|P^{(m)}_j(2g)|}{|j-m|}\le\left[
\sum_{j\ne m}|P_i^{(j)}(2g)|^2\right]^{1/2}\left[
\sum_{j\ne m}\frac{|P_j^{(m)}(2g)|^2}{|j-m|^2}\right]^{1/2}=\gamma_{i,m}\,
\sigma_m
$$

Due to the orthogonality of the transformation ${\bf U}(2g)$, as well as
above, we have
$$
\gamma_{i,m}=\sqrt{1-[P^{(m)}_i(2g)]^2}<1\,,\quad\mbox{for\,\,arbitrary\,\,$i$
\,\,and\,\,$m$}\,,
$$
and therefore
$$
\sum_{j\ne m}\frac{|P^{(j)}_i(2g)|\,|P^{(m)}_j(2g)|}{|j-m|}<\sigma_m=
\left[\sum_{j\ne m}\frac{|P_j^{(m)}(2g)|^2}{|j-m|^2}\right]^{1/2}
$$

Using the theorem on regular transformation, just as it was made for
$\lambda^{(2)}_m$, one can show that $\sigma_m\to0$ as $m\to\infty$.

Thus for the first term in~(\ref{23}), we have the unequality
$$
\sum_{i,j\ne m}\frac{
|P^{(i)}_m(2g)|\,|P^{(j)}_i(2g)|\,|P^{(m)}_j(2g)|}{|i-m|\,|j-m|}<\sigma_m\,
\sum_{i\ne m}\frac{
|P^{(i)}_m(2g)|}{|i-m|}
$$

Let us apply once again Cauchy's unequality to the sum on $i$ in the right
side of this unequality
$$
\sum_{i\ne m}\frac{|P^{(i)}_m(2g)|}{|i-m|}\le\left[
\sum_{i\ne m}|P_m^{(i)}(2g)|^2\right]^{1/2}\left[
\sum_{i\ne m}\frac{1}{(i-m)^2}\right]^{1/2}=\gamma_{m,m}\,f_m<f_m\,,
$$
$$
f_m=\left[\sum_{i\ne m}\frac{1}{(i-m)^2}\right]^{1/2}=\left[
\sum_{k=1}^m\frac{1}{k^2}+\sum_{k=1}^\infty\frac{1}{k^2}\right]^{1/2}<
\left[2\sum_{k=1}^\infty\frac{1}{k^2}\right]^{1/2}=\sqrt{\frac{\pi^2}{3}}=
\frac{\pi}{\sqrt3}
$$

It follows that
\begin{equation}
\label{*}
\sum_{i,j\ne m}\frac{
|P^{(i)}_m(2g)|\,|P^{(j)}_i(2g)|\,|P^{(m)}_j(2g)|}{|i-m|\,|j-m|}<
\frac{\pi}{\sqrt3}\,\sigma_m\,,
\end{equation}
and the unequality~(\ref{23}) takes the form
\begin{equation}
\label{24}
\left|\lambda_m^{(3)}\right|<\frac{(\omega_0/2)^3}{\omega^2}\,\sigma_m\,
\left[\frac{\pi}{\sqrt3}+
\left|1+(-1)^m\,P^{(m)}_m(2g)\right|\,\sigma_m\right]
\end{equation}

Since $\pi/\sqrt3\simeq1.81>1$ and $P^{(m)}_m(2g)\to0$ as $m\to\infty$ that
there exists such $m_0$ that
\begin{equation}
\label{25}
\left|1+(-1)^m\,P^{(m)}_m(2g)\right|<\pi/\sqrt3\,,\quad m>m_0
\end{equation}

Despite of that $|P^{(m)}_m(2g)|<1$ for arbitrary $m$, we could not prove that
the unequality~(\ref{25}) is valid for arbitrary $m$. Taking into account that
$\sigma_m<1$ and~(\ref{25}), we obtain from~(\ref{24})
\begin{equation}
\label{26}
\left|\lambda_m^{(3)}\right|<\frac{(\omega_0/2)^3}{3\,\omega^2}\,6\,
\frac{\pi}{\sqrt3}\,\sigma_m\,,\quad m>m_0\,;\quad \sigma_m\to0\,,\quad
m\to\infty
\end{equation}

Here, number $6$ is the number of components in the sum~(\ref{16}) for $k=3$.

Let us note that since $\sigma_m\to0$ as $m\to\infty$, we could write instead
of~(\ref{26}) more strong unequality, following from~(\ref{24}). But we shall
write just the unequality~(\ref{26}), following from~(\ref{25}), because just
this way can be used and for higher orders of a perturbation theory.

Let us consider the $k$-th order correction $\lambda_m^{(k)}$. Using Cauchy's
unequality and condition~(\ref{25}) one can show as well as above that the
absolute value of each term in the sum~(\ref{16}) is bounded by expression
\begin{equation}
\label{27}
\frac{(\omega_0/2)^k}{k\,\omega^{k-1}}\,\left(\frac{\pi}{\sqrt3}\right)^{k-2}
\,\sigma_m\,,\quad m>m_0\,,
\end{equation}
and hence
\begin{equation}
\label{28}
\left|\lambda_m^{(k)}\right|<
\frac{(\omega_0/2)^k}{k\,\omega^{k-1}}\,\left(\frac{\pi}{\sqrt3}\right)^{k-2}
\,N_k\,\sigma_m\,,\quad m>m_0\,,\quad k>2\,,
\end{equation}
where $N_k$ is the number of terms in the sum~(\ref{16}), i.e. the number of
solutions of the equation\\ $n_1+\ldots+n_k=k-1\,,\:\: n_i\ge0$
\begin{equation}
\label{29}
N_k=\frac{(2k-2)!}{[\,(k-1)!\,]^2}
\end{equation}

Let us show the estimation~(\ref{27}), for example, on the typical term,
entering in~(\ref{16}) at $k=4$
$$
h_m=\frac{(\omega_0)^4}{4\,\omega^3}\,\mbox{tr}
\left[{\bf P}\,{\bf S}_m^{0}\,{\bf P}\,{\bf S}_m^{0}\,{\bf P}\,
{\bf S}_m^{1}\,{\bf P}\,{\bf S}_m^{2}\right]=
\frac{(\omega_0)^4}{4\,\omega^3}\,P_{m,m}\,\sum_{i,j\ne m}
\frac{P_{m,i}\,P_{i,j}\,P_{j,m}}{(i-m)\,(j-m)^2}
$$
Using~(\ref{12}), we have
$$
|h_m|<\frac{(\omega_0/2)^4}{4\,\omega^3}\,
\left|1+(-1)^m\,P^{(m)}_m(2g)\right|\,
\sum_{i,j\ne m}\frac{|P_m^{(i)}(2g)|\,|P_i^{(j)}(2g)|\,|P_j^{(m)}(2g)|}
{|i-m|\,|j-m|}
$$
At last, using~(\ref{25}) and already obtained estimation~(\ref{*}), we obtain
$$
|h_m|<\frac{(\omega_0/2)^4}{4\,\omega^3}\,
\left(\frac{\pi}{\sqrt3}\right)^{2}\,\sigma_m\,,\quad m>m_0\,,
$$
that is the estimation~(\ref{27}) for $k=4$.

With the help of the unequality~(\ref{28}) we can estimate the reminder term
of the series~(\ref{15})
\begin{equation}
\label{30}
\left|\lambda_m-\sum_{k=0}^n\lambda_m^{(k)}\right|=
\left|\sum_{k=n}^\infty\lambda_m^{(k)}\right|<\frac{3\,\omega}{\pi^2}\,
\sigma_m\,\sum_{k=n}^\infty\frac{N_k}{k}\,\left(\frac{\omega_0\,\pi}
{\omega\,2\,\sqrt3}\,\right)^k\,,\quad m>m_0\,,\quad n>2
\end{equation}

From~(\ref{28}) it follows that asymptoticaly
$$
N_k=\frac{(2k-2)!}{[\,(k-1)!\,]^2}\sim\frac{4^k}{\sqrt k}\,,\quad k\to\infty
$$

It follows that the series in right part of~(\ref{30}) converges at
$\omega_0\le\omega\sqrt3/(2\pi)$. We can estimate it as follows. Since
$N_k<2^{2k-2}$, we have
$$
\left|\lambda_m-\sum_{k=0}^n\lambda_m^{(k)}\right|<\frac{3\,\omega}{4\pi^2}\,
\sigma_m\,\sum_{k=n}^\infty\left(\frac{2\,\omega_0\,\pi}
{\omega\,\sqrt3}\,\right)^k\,,\quad m>m_0\,,\quad n>2
$$
or
\begin{equation}
\label{31}
\left|\lambda_m-\sum_{k=0}^n\lambda_m^{(k)}\right|<\frac{3\,\omega}{4\pi^2}\,
\sigma_m\,\frac{\left(\frac{\textstyle2\,\omega_0\,\pi}
{\textstyle\omega\,\sqrt3}\right)^n}{1-
\frac{\textstyle2\,\omega_0\,\pi}{\textstyle\omega\,\sqrt3}}\,,
\quad m>m_0\,,\quad n>2
\end{equation}

Taking into account that $\sigma_m\to0$ as $m\to\infty$ and using~(\ref{31}),
~(\ref{22}),~(\ref{18}) and (\ref{19}), we obtain the following asymptotic
of eigenvalues $\lambda_m$
\begin{equation}
\label{32}
\lambda_m=m\omega+3\omega_0/2-g^2/\omega+o(1)\,,\quad m\to\infty
\end{equation}

Since the formulas~(\ref{11}),~(\ref{12}) differ only by sign, it easy to
see that the same asymptotic takes place and for eigenvalues
$\lambda_m^{(1)}$ of the operator ${\bf H}_1$. Thus, we have proved the
following result
\vspace{0.5cm}

\underline{\bf Theorem } :

If $\omega_0\le\omega\sqrt3/(2\pi)$, then the eigenvalues $\lambda_m^{(1)}$
and $\lambda_m^{(2)}$ of the operators ${\bf H}_1$ and ${\bf H}_2$ have the
asymptotic~(\ref{32}) and the reminder term of perturbation theory series
have the estimation~(\ref{31}).\\
---------------------------------------------------\\
\newpage

\bigskip
{\bf 4. Conclusion.}
\bigskip

In physical applications the main role plays not eigenvalues itself but a
difference of neighbouring eigenvalues, determining in the resonant case
$\omega_0=\omega$ the splitting of originally degenerate levels
$$
\Delta^{(1)}_m=\lambda_{2m+2}^{(1)}-\lambda_{2m+1}^{(1)}\,,\quad
m=0,1,2,\ldots
$$
$$
\Delta^{(2)}_m=\lambda_{2m+1}^{(2)}-\lambda_{2m}^{(2)}\,,\quad
m=0,1,2,\ldots
$$

From~(\ref{32}) it follows directly that
$$
\Delta^{(1,2)}_m\to\omega\,,\qquad m\to\infty
$$

It is in the sharp contradiction with the RWA. In the RWA an eigenvalues
appropriate, for example, to $\lambda_m^{(2)}$ (in resonant case
$\omega=\omega_0$), are defined by the expression
$$
\lambda_m=\omega\,(2m+2)\pm g\,\sqrt{2m+1}\,,\quad m=0,1,2,\ldots
$$

Therefore in the RWA the splitting grows as $\sqrt{2m}$.

This change of splitting undoubtedly should change the time dynamics of
quantum amplitudes, especially, when the average energy of a field mode is
sufficiently large.

We have proved the asymptotic formula~(\ref{32}) only at the condition
$\omega_0\le\omega\sqrt3/(2\pi)$. But the numerical calculations shows that
it is valid and for $\omega_0>\omega\sqrt3/(2\pi)$.

\bigskip
{\bf Acknowledgments.}
\bigskip

I am grateful to Prof. S.N. Naboko and Prof. N.M. Bogolubov for their
questions and useful remarks.


\begin{thebibliography}{12}

\bibitem{1}
Reik H. G., Nusser H., Amarante Ribeiro L. A., "Exact solution of
non-adiabatic model hamiltonians in solid state physics and optics.",
J. of Phys. A, v. 15, n. 11, 1982, p. 3491.
\bibitem{2}
Graham R., Hohnerbach M., "Quantum chaos of the two-level atom.",
Phys. Lett. A, v. 101, n. 2, 1984, p. 61.
\bibitem{3}
Kus M., Lewenstein M., "Exact isolated solutions for the class of quantum
optical systems.",  J. of Phys. A, v. 19, n. 2, 1986, p. 305.
\bibitem{4}
Lais P., Steimle T., "Squeezing in the Jaynes - Cummings model without
the RWA",  Optics communications, v. 78, n. 5,6 , 1990, p. 346.
\bibitem{5}
Feranchuk I. D., Komarov L. I., Ulyanenkov A. P., "Two - level system
in a one - mode quantum field : numerical solution on the basis of the
operator method.",  J. of  Phys. A, v. 29, 1996, p. 4035.
\bibitem{6}
Tur E.A., "Jaynes-Cummings model: Solution without rotating wave
approximation" , Optics and Spectroscopy, Vol. 89, n. 4, 2000, pp. 574-588.
\bibitem{7}
Tur E.A., "Energy Spectrum of the Hamiltonian of the Jaynes-Cummings Model
without Rotating-Wave Approximation" , Optics and Spectroscopy, Vol. 91, n. 6,
2001, pp. 899-902.
\bibitem{8}
Feynman R.P., Phys.Rev., {\bf 84}, 1951, 108.
\bibitem{9}
Schwinger J., Phys.Rev., {\bf 91}, 1953, 728.
\bibitem{10}
Kato T., Perturbation theory for linear operators, Springer-Verlag Berlin
$\cdot$Heidelberg$\cdot$New York, 1966.
\bibitem{11}
Szego G., "Orthogonal polynomials", New York, 1939.
\bibitem{12}
Hardy G., Divergent series, Oxford, 1949.

\end{thebibliography}
\end{document}